\documentclass[sensors,article,accept,moreauthors,pdftex,10pt,a4paper]{Definitions/mdpi}

\DeclareUnicodeCharacter{1E45}{\.{n}}
\DeclareUnicodeCharacter{1E41}{\.{m}}
\DeclareUnicodeCharacter{2003}{\quad}
\DeclareUnicodeCharacter{2009}{\thinspace}
\DeclareUnicodeCharacter{2002}{\enspace{}}
\DeclareUnicodeCharacter{2005}{\thinspace}
\DeclareUnicodeCharacter{0263}{\textipa{G}}
\DeclareUnicodeCharacter{A0}{~}
\DeclareUnicodeCharacter{2460}{\textcircled{\scriptsize{1}}}
\DeclareUnicodeCharacter{2461}{\textcircled{\scriptsize{2}}}
\DeclareUnicodeCharacter{2462}{\textcircled{\scriptsize{3}}}
\DeclareUnicodeCharacter{2463}{\textcircled{\scriptsize{4}}}
\DeclareUnicodeCharacter{2464}{\textcircled{\scriptsize{5}}}
\DeclareUnicodeCharacter{2465}{\textcircled{\scriptsize{6}}}
\DeclareUnicodeCharacter{2466}{\textcircled{\scriptsize{7}}}
\DeclareUnicodeCharacter{2467}{\textcircled{\scriptsize{8}}}
\DeclareUnicodeCharacter{2468}{\textcircled{\scriptsize{9}}}
\DeclareUnicodeCharacter{2070}{\textsuperscript{0}}
\DeclareUnicodeCharacter{2074}{\textsuperscript{4}}
\DeclareUnicodeCharacter{2075}{\textsuperscript{5}}
\DeclareUnicodeCharacter{2076}{\textsuperscript{6}}
\DeclareUnicodeCharacter{2077}{\textsuperscript{7}}
\DeclareUnicodeCharacter{2078}{\textsuperscript{8}}
\DeclareUnicodeCharacter{2079}{\textsuperscript{9}}
\DeclareUnicodeCharacter{02C2}{<}
\DeclareUnicodeCharacter{2033}{''}
\DeclareUnicodeCharacter{0229}{\c{e}}
\DeclareUnicodeCharacter{016F}{\r{u}}
\DeclareUnicodeCharacter{3AC}{\relax\ifmmode\acute{\alpha}\else $\acute{\alpha}$\fi}
\DeclareUnicodeCharacter{3AD}{\relax\ifmmode\acute{\varepsilon}\else $\acute{\varepsilon}$\fi}
\DeclareUnicodeCharacter{3AE}{\relax\ifmmode\acute{\eta}\else $\acute{\eta}$\fi}
\DeclareUnicodeCharacter{3AF}{\relax\ifmmode\acute{\iota}\else $\acute{\iota}$\fi}
\DeclareUnicodeCharacter{3CC}{\relax\ifmmode\acute{o}\else $\acute{o}$\fi}
\DeclareUnicodeCharacter{3CD}{\relax\ifmmode\acute{\upsilon}\else $\acute{\upsilon}$\fi}
\DeclareUnicodeCharacter{3CE}{\relax\ifmmode\acute{\omega}\else $\acute{\omega}$\fi}
\DeclareUnicodeCharacter{391}{A}
\DeclareUnicodeCharacter{392}{B}
\DeclareUnicodeCharacter{395}{E}
\DeclareUnicodeCharacter{396}{Z}
\DeclareUnicodeCharacter{397}{H}
\DeclareUnicodeCharacter{399}{I}
\DeclareUnicodeCharacter{39A}{K}
\DeclareUnicodeCharacter{39C}{M}
\DeclareUnicodeCharacter{39D}{N}
\DeclareUnicodeCharacter{39F}{O}
\DeclareUnicodeCharacter{3A1}{P}
\DeclareUnicodeCharacter{3A4}{T}
\DeclareUnicodeCharacter{3A7}{X}

\DeclareUnicodeCharacter{27E6}{\relax\ifmmode \llbracket \else $\llbracket$\fi}
\DeclareUnicodeCharacter{27E7}{\relax\ifmmode \rrbracket \else $\rrbracket$\fi}

\DeclareUnicodeCharacter{1D434}{\relax\ifmmode A \else $A$\fi}
\DeclareUnicodeCharacter{1D435}{\relax\ifmmode B \else $B$\fi}
\DeclareUnicodeCharacter{1D436}{\relax\ifmmode C \else $C$\fi}
\DeclareUnicodeCharacter{1D437}{\relax\ifmmode D \else $D$\fi}
\DeclareUnicodeCharacter{1D438}{\relax\ifmmode E \else $E$\fi}
\DeclareUnicodeCharacter{1D439}{\relax\ifmmode F \else $F$\fi}
\DeclareUnicodeCharacter{1D43A}{\relax\ifmmode G \else $G$\fi}
\DeclareUnicodeCharacter{1D43B}{\relax\ifmmode H \else $H$\fi}
\DeclareUnicodeCharacter{1D43C}{\relax\ifmmode I \else $I$\fi}
\DeclareUnicodeCharacter{1D43D}{\relax\ifmmode J \else $J$\fi}
\DeclareUnicodeCharacter{1D43E}{\relax\ifmmode K \else $K$\fi}
\DeclareUnicodeCharacter{1D43F}{\relax\ifmmode L \else $L$\fi}
\DeclareUnicodeCharacter{1D440}{\relax\ifmmode M \else $M$\fi}
\DeclareUnicodeCharacter{1D441}{\relax\ifmmode N \else $N$\fi}
\DeclareUnicodeCharacter{1D442}{\relax\ifmmode O \else $O$\fi}
\DeclareUnicodeCharacter{1D443}{\relax\ifmmode P \else $P$\fi}
\DeclareUnicodeCharacter{1D444}{\relax\ifmmode Q \else $Q$\fi}
\DeclareUnicodeCharacter{1D445}{\relax\ifmmode R \else $R$\fi}
\DeclareUnicodeCharacter{1D446}{\relax\ifmmode S \else $S$\fi}
\DeclareUnicodeCharacter{1D447}{\relax\ifmmode T \else $T$\fi}
\DeclareUnicodeCharacter{1D448}{\relax\ifmmode U \else $U$\fi}
\DeclareUnicodeCharacter{1D449}{\relax\ifmmode V \else $V$\fi}
\DeclareUnicodeCharacter{1D44A}{\relax\ifmmode W \else $W$\fi}
\DeclareUnicodeCharacter{1D44B}{\relax\ifmmode X \else $X$\fi}
\DeclareUnicodeCharacter{1D44C}{\relax\ifmmode Y \else $Y$\fi}
\DeclareUnicodeCharacter{1D44D}{\relax\ifmmode Z \else $Z$\fi}
\DeclareUnicodeCharacter{1D44E}{\relax\ifmmode a \else $a$\fi}
\DeclareUnicodeCharacter{1D44F}{\relax\ifmmode b \else $b$\fi}
\DeclareUnicodeCharacter{1D450}{\relax\ifmmode c \else $c$\fi}
\DeclareUnicodeCharacter{1D451}{\relax\ifmmode d \else $d$\fi}
\DeclareUnicodeCharacter{1D452}{\relax\ifmmode e \else $e$\fi}
\DeclareUnicodeCharacter{1D453}{\relax\ifmmode f \else $f$\fi}
\DeclareUnicodeCharacter{1D454}{\relax\ifmmode g \else $g$\fi}
\DeclareUnicodeCharacter{1D456}{\relax\ifmmode i \else $i$\fi}
\DeclareUnicodeCharacter{1D457}{\relax\ifmmode j \else $j$\fi}
\DeclareUnicodeCharacter{1D458}{\relax\ifmmode k \else $k$\fi}
\DeclareUnicodeCharacter{1D459}{\relax\ifmmode l \else $l$\fi}
\DeclareUnicodeCharacter{1D45A}{\relax\ifmmode m \else $m$\fi}
\DeclareUnicodeCharacter{1D45B}{\relax\ifmmode n \else $n$\fi}
\DeclareUnicodeCharacter{1D45C}{\relax\ifmmode o \else $o$\fi}
\DeclareUnicodeCharacter{1D45D}{\relax\ifmmode p \else $p$\fi}
\DeclareUnicodeCharacter{1D45E}{\relax\ifmmode q \else $q$\fi}
\DeclareUnicodeCharacter{1D45F}{\relax\ifmmode r \else $r$\fi}
\DeclareUnicodeCharacter{1D460}{\relax\ifmmode s \else $s$\fi}
\DeclareUnicodeCharacter{1D461}{\relax\ifmmode t \else $t$\fi}
\DeclareUnicodeCharacter{1D462}{\relax\ifmmode u \else $u$\fi}
\DeclareUnicodeCharacter{1D463}{\relax\ifmmode v \else $v$\fi}
\DeclareUnicodeCharacter{1D464}{\relax\ifmmode w \else $w$\fi}
\DeclareUnicodeCharacter{1D465}{\relax\ifmmode x \else $x$\fi}
\DeclareUnicodeCharacter{1D466}{\relax\ifmmode y \else $y$\fi}
\DeclareUnicodeCharacter{1D467}{\relax\ifmmode z \else $z$\fi}

\usepackage{xcolor, colortbl}
\usepackage[normalem]{ulem}
\usepackage{longtable}
\usepackage{threeparttable}
\usepackage{booktabs,multirow}
\usepackage{attrib}
\usepackage{epstopdf}
\usepackage{subfigure}
\usepackage{mathcomp}
\usepackage{amsfonts}
\usepackage{amssymb}
\usepackage{CJKutf8}
\usepackage{pxfonts}
\makeatletter
\def\T@n@@nc@d@ngM@cr@M@d{}
\def\LY@n@@nc@d@ngM@cr@M@d{}
\makeatother

\usepackage{fancyvrb}
\usepackage{pifont}
\usepackage{setspace}

\graphicspath{{./Definitions/}}

\usepackage{upgreek}
\usepackage{wasysym}

\setitemize{parsep=6pt,itemsep=0pt,leftmargin=*,labelsep=5.5mm,align=parleft}
\setenumerate{parsep=6pt,itemsep=0pt,leftmargin=*,labelsep=5.5mm,align=parleft}
\setlist[description]{itemsep=0mm}

\let\originalleft\left
\let\originalright\right
\renewcommand{\left}{\mathopen{}\mathclose\bgroup\originalleft}
\renewcommand{\right}{\aftergroup\egroup\originalright}

\let\orignewcommand\newcommand  % store the original \newcommand
\let\newcommand\providecommand  % make \newcommand behave like \providecommand
\usepackage{verse}
\let\newcommand\orignewcommand  % use the original `\newcommand` in future



\newsavebox\foobox

\renewcommand{\dagger}{\mathchar"2279}
\renewcommand{\ddagger}{\mathchar"227A}
\usepackage{bm}

\usepackage{seqsplit}
  \newlength{\cellWidtha}
  \newlength{\cellWidthb}
  \newlength{\cellWidthc}
  \newlength{\cellWidthd}
  \newlength{\cellWidthe}
  \newlength{\cellWidthf}

\newcommand{\fig}[1]{Figure~\ref{#1}}
\newcommand{\sect}[1]{Section~\ref{#1}}
\newcommand{\tabref}[1]{Table~\ref{#1}}

\usepackage{cleveref}
\crefname{figure}{Figure}{Figures}
\crefname{table}{Table}{Tables}
\crefrangelabelformat{figure}{#3#1#4--#5#2#6}
\crefrangelabelformat{table}{#3#1#4--#5#2#6}
\crefname{section}{Section}{Sections}
\crefname{paragraph}{Section}{Sections}
\crefrangelabelformat{section}{#3#1#4--#5#2#6}
\crefname{appendix}{Appendix}{Appendices}
\crefrangelabelformat{appendix}{#3#1#4--#5#2#6}
\crefname{scheme}{Scheme}{Schemes}
\crefrangelabelformat{scheme}{#3#1#4--#5\crefstripprefix{#1}{#2}#6}
\crefname{chart}{Chart}{Charts}
\crefrangelabelformat{chart}{#3#1#4--#5\crefstripprefix{#1}{#2}#6}

\newcommand{\PreserveBackslash}[1]{\let\temp=\\#1\let\\=\temp}

\usepackage{tikz}

\makeatletter
\g@addto@macro{\UrlBreaks}{\UrlOrds}
\makeatother

\makeatletter
\let\org@underset\underset
\renewcommand{\underset}[2]{
  \ifthenelse{\equal{#1}{\}\ }}{\underbrace{#2}}{\org@underset{#1}{#2}}
}

\newcommand{\mmathit}[1]{
  \ifthenelse{\equal{#1}{\ln}}{\mathit{ln}}{
    \ifthenelse{\equal{#1}{\max}}{\mathit{max}}{\mathit{#1}}
  }
}

\makeatother
\robustify{\footnote}

\firstpage{1}

  \articlenumber{6723}

\pubvolume{20}
\issuenum{23}
\pubyear{2020}
\copyrightyear{2020}

\history{Received: 6 October 2020; Accepted: 19 November 2020; Published: 24 November 2020}

\updates{yes} % If there is an update available, un-comment this line

\usepackage[russian,english]{babel}

\usepackage[T5,T1]{fontenc}

\renewcommand{\mathit}[1]{#1}
\renewcommand{\mathsf}[1]{#1}

\Title{Wide Field Spectral Imaging with Shifted Excitation Raman Difference Spectroscopy Using the Nod and Shuffle~Technique}

\Author{Florian~Korinth~\textsuperscript{1}, Elmar~Schmälzlin~\textsuperscript{2}, Clara~Stiebing~\textsuperscript{1}, Tanya~Urrutia~\textsuperscript{2}\href{https://orcid.org/0000-0001-6746-9936}{\orcidicon}, Genoveva~Micheva~\textsuperscript{2}, Christer~Sandin~\textsuperscript{3}, Andr{\fontencoding{T5}\selectfont{\'e}}~Müller~\textsuperscript{4}, Martin~Maiwald~\textsuperscript{4}\href{https://orcid.org/0000-0003-1166-5529}{\orcidicon}, Bernd~Sumpf~\textsuperscript{4}\href{https://orcid.org/0000-0001-5044-955X}{\orcidicon}, Christoph~Krafft~\textsuperscript{1}\textsuperscript{,}*\href{https://orcid.org/0000-0003-1049-0560}{\orcidicon}, Günther~Tränkle~\textsuperscript{4}, Martin~M.~Roth~\textsuperscript{2}\href{https://orcid.org/0000-0003-2451-739X}{\orcidicon} and Jürgen~Popp~\textsuperscript{1}\textsuperscript{,}\textsuperscript{5}\href{https://orcid.org/0000-0003-4257-593X}{\orcidicon}}
\AuthorNames{Florian Korinth, Elmar Schmälzlin, Clara Stiebing, Tanya Urrutia, Genoveva Micheva, Christer Sandin, André Müller, Martin Maiwald, Bernd Sumpf, Christoph Krafft, Günther Tränkle, Martin M. Roth and Jürgen Popp}
\address{\textsuperscript{1} \quad Leibniz Institute of Photonic Technology~(Leibniz IPHT), Research Alliance “Health~Technologies”, Albert-Einstein-Stra{\ss}e 9, 07743 Jena, Germany; florian.korinth@leibniz-ipht.de~(F.K.); clara.stiebing@leibniz-ipht.de~(C.S.); juergen.popp@leibniz-ipht.de~(J.P.)

\textsuperscript{2} \quad Leibniz Institute for Astrophysics Potsdam~(AIP), Research Alliance “Health~Technologies”, An der Sternwarte 16, 14482 Potsdam, Germany; eschmaelzlin@aip.de~(E.S.); turrutia@aip.de~(T.U.); gmicheva@aip.de~(G.M.); mmroth@aip.de~(M.M.R.)

\textsuperscript{3} \quad Sandin Advanced Visualization, Tylögränd 14, 12156 Johanneshov, Sweden; christersandin@yahoo.se

\textsuperscript{4} \quad Ferdinand-Braun-Institut, Leibniz-Institut für Höchstfrequenztechnik, Research Alliance “Health~Technologies”, Gustav-Kirchhoff-Str. 4, 12489 Berlin, Germany; andre.mueller@fbh-berlin.de~(A.M.); martin.maiwald@fbh-Berlin.de~(M.M.); bernd.sumpf@fbh-berlin.de~(B.S.); guenther.traenkle@fbh-berlin.de~(G.T.)

\textsuperscript{5} \quad Institute of Physical Chemistry and Abbe Center of Photonics, Friedrich Schiller University Jena, Helmholtzweg 4, 07743 Jena, Germany}

\corres{Correspondence: christoph.krafft@leibniz-ipht.de}

\abstract{Wide field Raman imaging using the integral field spectroscopy approach was used as a fast, one shot imaging method for the simultaneous collection of all spectra composing a Raman image. For the suppression of autofluorescence and background signals such as room light, shifted~excitation Raman difference spectroscopy (SERDS) was applied to remove background artifacts in Raman spectra. To reduce acquisition times in wide field SERDS imaging, we adapted the nod and shuffle technique from astrophysics and implemented it into a wide field SERDS imaging setup. In our adapted version, the nod corresponds to the change in excitation wavelength, whereas the shuffle corresponds to the shifting of charges up and down on a Charge-Coupled Device (CCD) chip synchronous to the change in excitation wavelength. We coupled this improved wide field SERDS imaging setup to diode lasers with 784.4/785.5 and 457.7/458.9 nm excitation and applied it to samples such as paracetamol and aspirin tablets, polystyrene and polymethyl methacrylate beads, as well as pork meat using multiple accumulations with acquisition times in the range of 50 to 200 ms.~The results tackle two main challenges of SERDS imaging: gradual photobleaching changes the autofluorescence background, and multiple readouts of CCD detector prolong the acquisition~time.}

\keyword{Raman spectroscopy; SERDS; wide field imaging; nod and shuffle; photobleaching; autofluorescence; integral field spectroscopy; background suppression; diode~lasers}

\makeatletter
\DeclareRobustCommand*\textsubscript[1]{%
  \@textsubscript{\selectfont#1}}
\def\@textsubscript#1{%
  {\m@th\ensuremath{_{\mbox{\fontsize\sf@size\z@#1}}}}}
\makeatother

\usepackage{sansmath}

\begin{document}
\section{Introduction \label{sect:sec1-sensors-974262}}

Raman spectroscopy provides a label-free, non-destructive insight into the biochemical composition of a sample.~It can be used for the pathological assessment of biological samples such as cells and tissues~\cite{B1-sensors-974262,B2-sensors-974262,B3-sensors-974262,B4-sensors-974262,B5-sensors-974262}. Although the majority of biomolecules such as nucleic acids, proteins, lipids, and carbohydrates do not fluoresce at visible excitation, autofluorescence is often excited along with Raman scattering in biological samples~\cite{B6-sensors-974262}. Since the fluorescence absorption cross-section is much larger than the Raman scattering cross-section, even trace amounts of autofluorescent molecules often result in high-intensity spectral backgrounds. This high background can mask the Raman fingerprint information and increases the shot noise in spectra~\cite{B7-sensors-974262}.

Approaches to solve this challenge can be divided into computational and instrumental methods~\cite{B8-sensors-974262}.~Computational methods correct Raman spectra for their background using mathematical algorithms such as extended multiplicative scatter correction (EMSC)~\cite{B9-sensors-974262,B10-sensors-974262}, rubberband~\cite{B11-sensors-974262,B12-sensors-974262}, sensitive~nonlinear iterative peak (SNIP)~\cite{B13-sensors-974262,B14-sensors-974262}, and polynomial fittings~\cite{B15-sensors-974262}.~Examples for the application of polynomial fitting are the correction during \mbox{in vivo} Raman studies of colorectal tissue~\cite{B16-sensors-974262} and brain cancer~\cite{B17-sensors-974262}. EMSC was used for the autofluorescence correction of Raman spectra of bladder biopsies~\cite{B18-sensors-974262}.~Depending on the complexity of the algorithm, the computational data processing can be time~consuming.

Instrumental methods for autofluorescence suppression enable less time consuming data processing, but they usually require more complex experimental setups.~Examples for these methods are time-gated methods~\cite{B19-sensors-974262,B20-sensors-974262}, phase or wavelength modulation techniques~\cite{B21-sensors-974262,B22-sensors-974262,B23-sensors-974262,B24-sensors-974262}, and shifted excitation Raman difference spectroscopy (SERDS)~\cite{B25-sensors-974262}. For SERDS, two Raman spectra are measured at the same spatial position with two slightly shifted excitation wavelengths. Raman bands, which~are a function of the exciting laser light, follow the shift in excitation wavelength. The background, a~function of the instrument (filters, optics, and detection system), the sample’s autofluorescence, room~light, and~other background components, does not shift. The two spectra are subtracted from each other, ideally~resulting in a background-free difference spectrum with a first derivative-like appearance. For~a proper reconstruction of the Raman spectrum from the difference spectrum, the shift in excitation wavelength, and therefore also the shift in relative wavenumbers, should be selected in accordance with the band widths of the expected Raman spectra. The target classification can be derived directly from the SERDS spectra, which was demonstrated for pollen~\cite{B26-sensors-974262}, or from algorithm-based reconstructed Raman spectra, which was demonstrated for formalin fixed patient samples of breast tumors~\cite{B27-sensors-974262}. Further studies have already been published on biological samples using the SERDS technique. The~investigated samples were molar tooth slices and \mbox{in vivo} human skin~\cite{B28-sensors-974262}, algae~produced complex polysaccharides~\cite{B29-sensors-974262}, tissues from different animals~\cite{B30-sensors-974262,B31-sensors-974262,B32-sensors-974262}, ex vivo oral squamous cell carcinoma biopsies~\cite{B33-sensors-974262}, and spores of Aspergillus nidulans conidia~\cite{B34-sensors-974262}.

There are only few examples of SERDS imaging so far, in which each point of the image corresponds to a difference spectrum. In the first implementation, such images were collected by performing a point-by-point raster scan of the sample, measuring each point twice with two different wavelengths. Cordero et al.~\cite{B7-sensors-974262} measured a beef steak in this fashion.~After clustering the mean spectra of each cluster and assigning them to two different types of proteins and lipid, they compared the results of different wavelength shifts for SERDS with results based on EMSC and the 1st derivative of the EMSC corrected spectra.~A second implementation applied a so-called integral field spectroscopy (IFS) approach~\cite{B35-sensors-974262,B36-sensors-974262}, which belongs to the wide field imaging techniques. Whereas point scanning or line mapping approaches move the laser spot or line across the sample to sequentially collect a Raman image, wide field imaging instruments project the sample onto a multichannel detector such as a Charge-Coupled Device (CCD) chip and acquire the full Raman image simultaneously~\cite{B37-sensors-974262}. The~spectral information can be obtained by fixed wavelength or tunable filters. In the context of light sheet Raman imaging, an interferometric detection scheme was developed~\cite{B38-sensors-974262}. More than 1000 Raman images were registered within minutes from which the Raman spectra were calculated via Fourier transformation. Another group of Raman imaging techniques generates several laser foci spots on the sample simultaneously by using fiber arrays, micro lens arrays, galvomirrors, or spatial modulators (such as liquid-crystal spatial modulators or digital micromirror devices). Then, the spectra from all foci are imaged on a CCD either by the use of integral field spectrographs or by producing several pinholes using the spatial modulators~\cite{B39-sensors-974262}.

In astronomical observatories, integral field spectroscopy (IFS)~\cite{B40-sensors-974262} is used to create a complete spectral image of cosmic objects in one shot without raster scanning the object, which saves valuable and rare observation time. For this one shot imaging, the image is optically dissected by a so-called integral field unit (IFU). The sections (e.g., slices, spots, squares, hexagons) are re-arranged in front of the rather large entrance slit of a wide field spectrograph or even several such spectrographs. One~of the options to realize an IFU is using a fiber bundle that is arranged in a two-dimensional array on the side facing the sample and in a row-like fashion on the side facing the spectrograph. The row of fiber front surfaces acts as long pseudo-slit at the entrance side of the spectrograph. If the fibers of the bundle have round cores, the image will be dissected into round spots. At the spectrograph, the dispersed signals emerging from the fibers are collected by a large area detector. Each signal trace correlates to a specific image location. The spectrograph is designed in a way that signal traces originating from light at the ends of the entrance slit show nearly the same quality as light traces originating from the slit’s center, meaning the spectrographs works over a “wide field” with regard to signal position at the entrance. The raw signal is processed into a data cube that contains the full spectral and lateral information. So~far, the most powerful IFS spectrograph developed for astronomy is M.U.S.E. (Multi~Unit Spectroscopic Explorer)~\cite{B41-sensors-974262} at the Very Large Telescope Observatory, Paranal, Chile. It~consists of 24 IFUs, each~with its own attached spectrograph, and it detects a total of 90,000~spectra (and~therefore image pixels called “spaxels”) with a spectral resolution of 0.25 nm in the range of 465 to 930 nm in one exposure simultaneously. An IFS spectrograph with 400 spaxels, built as a clone of a M.U.S.E. spectrograph module, was successfully used for Raman and SERDS imaging~\cite{B35-sensors-974262,B36-sensors-974262}.~A custom-made probe head focuses the excitation laser in 400 measuring spots on the sample and guides the backscattered Raman signal into the IFS spectrograph. This setup was used by Schmälzlin et al.~\cite{B36-sensors-974262} for Raman and SERDS imaging of different samples, e.g., a cross-section of a pig ear, skin, and a dissolving brown sugar~cube.

Another technique from astrophysics that could be translated to wide field SERDS imaging is the nod and shuffle concept~\cite{B42-sensors-974262,B43-sensors-974262,B44-sensors-974262}.~This concept was originally developed to measure deep spectra of very faint galaxies whose surface brightness is usually a fraction of the night sky brightness, sometimes~orders of magnitude fainter than the sky. In order to overcome the systematic errors introduced by atmospheric background intensity fluctuations, it is necessary to employ a technique called beam switching, which essentially means to measure pairs of (object + sky) and (sky) in rapid succession. While fast readout detectors such as infrared arrays have no problem to accomplish this in real time, Charge-Coupled Devices (CCD) in the optical wavelength domain incur substantial readout times per image. To accommodate beam switching with CCDs, the original nod and shuffle technique allows for a “nod” as a small offset of the telescope to switch its focus between the target and the neighboring background night sky. The “shuffle” stands for shifting charges on a frame transfer CCD from an active exposure area into one of two storage areas: one for (object + sky), another one for (sky). By synchronizing the nod process with the charge shuffle on the chip, it is possible to create the beam switching short exposure pairs that are necessary to eliminate the rapid sky brightness fluctuations without having to read out the CCD every time. After a sufficiently high total exposure time has been acquired by a series of many nod--shuffles (of order several tens), the exposure is stopped, and~the entire CCD is read out. This technique works particularly well with fiber-coupled spectrographs, where the spacing of fibers on the entrance slit provides gaps between spectral traces on the detector and the gaps are wide enough to accommodate an extra trace between two adjacent spectral traces in an interlaced fashion.~A virtue of this technique lies in the ability to subtract (object + sky) and (sky) on a pixel-by-pixel basis, thus avoiding further systematic errors that occur in the process of data reduction, specifically optimal extraction of the spectral traces: the illumination of each pair of pixels in the (object + sky) and (sky) storage areas has occurred under (almost) exactly the same~circumstances.

In this work, the nod and shuffle technique from astrophysics is adapted and implemented into a wide field SERDS imaging setup. As a result, this concept addresses the two main challenges of SERDS imaging.~First, photobleaching between consecutive Raman spectra at two excitation wavelengths reduces the auto-fluorescent background and causes a residual background slope in the difference spectrum.~Second, short exposure times in the millisecond range can be applied to minimize the photobleaching effect that results in low signal to noise ratios or long total acquisition times for hundreds of single spectra.~Since the CCD chip has to be read out in a traditional setup after each acquisition and readout time can take up to a minute or more in particular for large CCDs, the~strategy of very short acquisitions and multiple readouts would increase the overall measurement time tremendously. The~nod and shuffle technique can be adapted to support SERDS imaging and overcome these challenges of SERDS imaging. The nod corresponds to the shift in excitation wavelength at the same spatial position, whereas the shuffle remains the shuffling of the charges on the CCD, saving, and accumulating both signals on the same CCD until numerous cycles revealed to sufficiently high signal to noise ratio. For a single channel setup with a rectangle CCD chip (1024 $\times$ 256 pixels), this was previously demonstrated by Sowoidnich et al.~\cite{B45-sensors-974262,B46-sensors-974262} as a Charge-Shifting Charge-Coupled Device, based on the previous works of Heming et al.~\cite{B47-sensors-974262}.

In comparison to a previous conference proceeding paper~\cite{B48-sensors-974262}, this publication provides more details of the nod and shuffle approach and the laser diode development, and it demonstrates wide field spectral imaging with nod and shuffle SERDS for a tissue specimen for the first~time.

\section{Materials and Methods \label{sect:sec2-sensors-974262}}
\unskip
\subsection{Wide Field SERDS Setup \label{sect:sec2dot1-sensors-974262}}

In the experiments, an IFS spectrograph unit of the M.U.S.E. system (Winlight System, Pertuis, France) with a large area 4096 $\times$ 4096 CCD chip (CCD231, Teledyne e2v, UK) was used. A detailed description of the spectrograph and detector system can be found in~\cite{B49-sensors-974262} and~\cite{B50-sensors-974262}. Two different laser sources were used during the presented experiments. They were selected for their shift in their emitted wavelength, which should correlate to the expected band widths of the Raman profiles of the samples (see below in \tabref{tabref:sensors-974262-t001}).

\begin{enumerate}[label=\arabic*.]
\item  A fiber-coupled near infrared (NIR) laser source LS 2-VBG, Ushio, Tokyo, Japan (formerly PD-LD, Pennington, USA) was used. The laser emits at 784.43 and 785.48 nm with a maximum laser power of 400 mW. This 1.05 nm shift (17 cm\textsuperscript{$-$1}) in excitation wavelength was the largest possible shift without a high loss of laser intensity due to the filters used in the setup. This small shift in wavenumbers was used for the investigation of two pharmaceutical pills, which show Raman spectra with narrow band~profiles.
\item  A custom-made laser source based on two external wavelength stabilized blue laser diodes with spatially overlapping laser emissions at 457.74 and 458.90 nm in a conduction cooled package mount with a 25 $\times$ 25 mm\textsuperscript{2} footprint was designed and realized~\cite{B51-sensors-974262}.~The diode lasers are wavelength stabilized at both target wavelengths with volume Bragg gratings with spectral bandwidths of \textless{}0.2 nm (\textless{}10 cm\textsuperscript{$-$1}) and diffraction efficiencies of 15\%. The spectral distance of 1.16 nm (55 cm\textsuperscript{$-$1}) is selected for SERDS with respect to the spectral resolution of the spectrometer with 0.22 nm ($\approx$10.4 cm\textsuperscript{$-$1}) and for the investigation of biological tissue samples, since these samples have broader Raman band structures. For the spatial overlap of both laser emissions, using a high-reflection coated prism as well as a polarizing beam splitter, one laser is rotated by 90$^\circ$. During the measurements, the laser was mounted on a heatsink (hsa-series, Ostech, Berlin, Germany), which was set to an operating temperature of 40 $^\circ$C. For the individual operation of both laser diodes, two laser drivers were used (ds-series, Ostech, Berlin, Germany). The~optical output power at both wavelengths was 0.4 W at 0.5 A. \fig{fig:sensors-974262-f001} shows the corresponding emission~spectra.
\begin{figure}[H]
\centering
\includegraphics[scale=1.65]{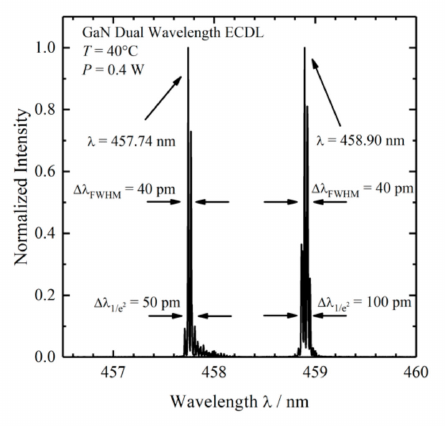}
\caption{Emission spectra of the dual-wavelength GaN external cavity diode laser (ECDL) measured at an operating temperature of 40$^\circ$C and an optical output power of 0.4~W.}
\label{fig:sensors-974262-f001}
\end{figure}

\end{enumerate}

\begin{figure}[H]
\centering
\includegraphics[scale=2.3]{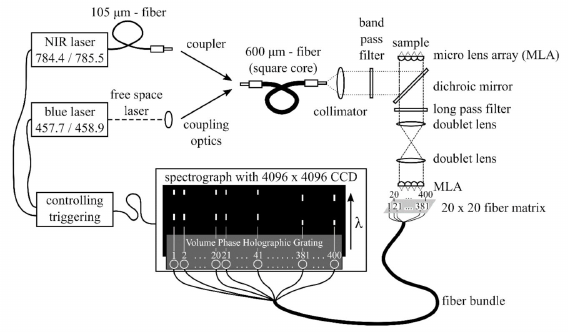}
\caption{Wide field shifted excitation Raman difference spectroscopy (SERDS) imaging setup based on two micro lens~arrays.}
\label{fig:sensors-974262-f002}
\end{figure}

The basic setup for Raman measurements was already reported in~\cite{B36-sensors-974262}. Briefly, 400 fibers of a bundle are arranged in a line at the spectrograph entrance to function as an IFU. On the sample side of the fiber bundle, the fibers form a 1 cm $\times$ 1 cm square, consisting of a 20 $\times$ 20 fiber matrix with a micro lens array (MLA) attached, so that the backscattered Raman signal can be efficiently coupled into the fibers. The laser light is guided via mirrors and filters on the sample. Two different probe head configurations were used: In the first configuration, another MLA focuses the laser light onto the sample by creating 400 foci (50 $\upmu$m diameter; 0.5 mm center to center distance between focal spots) simultaneously while also collecting the Raman signal. This results in an imaging area of 1 cm\textsuperscript{2}. The~probe head in combination with the sample-side MLA was originally designed to measure thick skin samples with a low depth resolution. The low confocality is a consequence of 1 to 1 projection of 50 $\upmu$m diameter foci onto 114 $\upmu$m fiber cores to compensate alignment tolerances~\cite{B52-sensors-974262}. Due to the scattering, the achievable resolution was only 0.5 mm and, thus, a denser packing of the fibers was not useful. However, a far denser packing is possible and was already realized~\cite{B53-sensors-974262}. \fig{fig:sensors-974262-f002} shows a schematic of the setup.~The doublet lenses act as a telescope.~They project the front surface of the sample onto the front surface of the fiber array. Furthermore, they project the exit pupil of the sample-side lens array to the entrance pupil of the fiber-side lens array. The second configuration with a field of view of approximately 0.02 cm\textsuperscript{2} is used for smaller samples. Here, the MLA on the sample side is exchanged by an objective lens (Magnification = 10$\times$, NA = 0.25). The two 50 mm doublet lenses are exchanged by a 200 mm tube lens. To implement the nod and shuffle technique, the wavelength axis has to be parallel to the readout register of the spectrograph, since the charge shuffle is only possible in the direction of the readout registers. Therefore the wavelength axis has to be orthogonal to the readout direction. In~the original configuration of the spectrograph, this was not the case~\cite{B36-sensors-974262}; thus, the CCD chip had to be rotated by 90$^\circ$. Unfortunately, a rotation of the existing CCD chip was not possible, since~it had a graded index anti-reflection coating that was designed to maximize the throughput along the CCD chip’s original wavelength axis. After the rotation, the~profile of the anti-reflection coating would not match the wavelengths of the incident light anymore. Thus, a new CCD chip of the same type but with a uniform anti-reflection coating was purchased and built-in in the proper rotation. The~loss of transmission caused by the simpler coating was insignificant. The camera control was changed to enable the shifting of the charges and the triggering of the lasers according to the settings in the user interface. For the generation of data cubes with single Raman spectra and difference spectra, the data reduction software also had to be adjusted to the new detector raw signal, which consisted of twice as many signal traces after the nod and shuffle~changes.

\subsection{Nod and Shuffle \label{sect:sec2dot2-sensors-974262}}

\begin{figure}[H]
\centering
\includegraphics[scale=2.3]{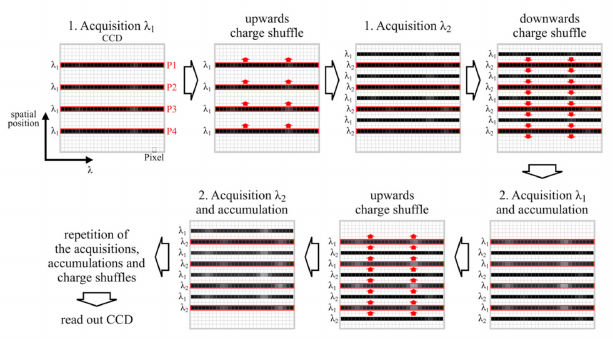}
\caption{Illustration of the nod and shuffle technique by schematically showing the Charge-Coupled Device (CCD) chip (one square denotes one pixel on the CCD) and the shifting of charges (red arrows) after each acquisition and accumulation. The black lines symbolize the spectral signal traces with the corresponding excitation wavelengths to the left of them. The white spots inside the black lines display increasing signal strength, i.e., Raman bands that get more pronounced during the accumulation. The~red boxes show the respective regions on the CCD, which are illuminated during the acquisition, where each box corresponds to the same spaxel (spatial Position P1, P2, P3, and P4) of the Raman image. For~simplicity, only four spaxels of the Raman image are~shown.}
\label{fig:sensors-974262-f003}
\end{figure}

As briefly explained in the introduction the nod and shuffle technique for SERDS is a technique adapted from astrophysics. In SERDS, the nod corresponds to the change in excitation wavelength and the shuffle is, as in astronomy, the shift of charges on the CCD chip. In \fig{fig:sensors-974262-f003}, the interlaced nod and shuffle technique for SERDS is~depicted.

At first, a wide field Raman image of the sample is measured with the first excitation wavelength ($\uplambda$\textsubscript{1}). Each signal trace corresponds to a full spectrum at a distinct spatial position. Then, all charges are shifted up by 5 pixels on the CCD, the excitation wavelength is changed to $\uplambda$\textsubscript{2}, and a new wide field Raman image of the sample is taken.~A CCD consists of a light-sensitive area organized in vertical columns of electrodes. Four electrodes in series form a pixel in our case. They are separated horizontally by potential wells. Applying high or low voltage to the electrodes in a special scheme allows collecting electrons (charge) underneath.~Clocking the electrodes with a dedicated voltage pattern moves the charge from electrode to electrode. All electrodes are equivalent and therefore charge transfer vertically up and down is possible.~A shift of 5 pixels transfers the signal traces into the centers of the spaces in between their recording positions.~A transfer to the centers minimizes the risk of crosstalk to the respective new recorded upper and lower signal traces. Then, all charges are shifted down by 5 pixels on the CCD, and the excitation wavelength is shifted back to $\uplambda$\textsubscript{1}. This ends the first acquisition cycle. By repeating these steps, several acquisitions with two different wavelengths can be accumulated without reading out the CCD chip in between measurements. The challenge of this technique is the high instrumental complexity and, in case of very intense signals, the danger of crosstalk due to the lower pixel distance between the signal traces. However, there are several advantages to this method:\begin{enumerate}[label=\arabic*.]
\item  Readout noise is only applied once to the complete set of~data.
\item  Readout time only factors into the complete measurement time once. As for a large CCD chip, the readout time can take up to a minute or more, the total measurement time is significantly increased, when reading out the CCD chip after every single measurement. In our case, a sample was measured with 200 accumulations for each wavelength before reading out the CCD once. If the CCD was read out after each acquisition, readout times alone would have taken 400 min instead of 1 min with nod and~shuffle.
\item  The technique allows measurements with very short acquisition times, which were accumulated with multiple iterations over a long time period using a wide field imaging setup. The~accumulation of single measurements with very short acquisition times is favorable in the presence of signal fluctuations, e.g., photobleaching and fluctuating high-intensity background light (see~\sect{sect:sec3dot2dot1-sensors-974262}).
\item  Spectra measured at the same spatial position with different wavelengths are recorded on the same pixels of the CCD chip. Therefore, Raman images of both excitation wavelengths have the same artifacts originating from the sensitivity and noise variations of the CCD~pixels.
\end{enumerate}

\subsection{Samples and Experimental Parameters \label{sect:sec2dot3-sensors-974262}}

The samples used for the experiments were:\begin{enumerate}[label=$\bullet$]
\item Sample 1: Paracetamol tablet and a piece of an aspirin tablet were arranged on the large field of view probe head. Both samples show intense Raman~bands.
\item Sample 2: Polystyrene (PS) beads (50 $\upmu$m diameter) and polymethyl methacrylate (PMMA) beads (120 $\upmu$m diameter) were mixed and applied on a CaF\textsubscript{2} window. Both samples show intense Raman bands and have diameters to demonstrate the lateral~resolution.
\item Sample 3:~The lipid-rich part of a pork chop fresh from a butcher was placed on a CaF\textsubscript{2} window. Pork tissue served as model for biological material with less intense Raman bands and elevated~background.
\item Sample 4: The transition region from meat to fat of a pork chop fresh from a butcher was placed on a CaF\textsubscript{2}~window.
\end{enumerate}

The technical measurement conditions are shown in \tabref{tabref:sensors-974262-t001}. The maximum acquisition time of 2 s for no. 3 and 4 gave the highest overall signal intensity on the CCD chip without crosstalk. This time was divided by the shortest achievable acquisition time of 50 ms for the laser system and resulted in 40~acquisitions. In the future, control of the lasers and the charge shift shuffling are improved to realize even shorter acquisitions and systematic studies are planned to obtain a good signal-to-noise ratio and zero baseline at the same~time.
    \begin{table}[H]
    \tablesize{\small}
    \centering
    \caption{Overview over the technical measurement conditions shown by~samples.}
    \label{tabref:sensors-974262-t001}

\setlength{\cellWidtha}{\textwidth/6-2\tabcolsep-0.65in}
\setlength{\cellWidthb}{\textwidth/6-2\tabcolsep+.5in}
\setlength{\cellWidthc}{\textwidth/6-2\tabcolsep+.25in}
\setlength{\cellWidthd}{\textwidth/6-2\tabcolsep-0.25in}
\setlength{\cellWidthe}{\textwidth/6-2\tabcolsep-0in}
\setlength{\cellWidthf}{\textwidth/6-2\tabcolsep-0in}
\scalebox{1}[1]{\begin{tabular}{>{\PreserveBackslash\centering}m{\cellWidtha}>{\PreserveBackslash\centering}m{\cellWidthb}>{\PreserveBackslash\centering}m{\cellWidthc}>{\PreserveBackslash\centering}m{\cellWidthd}>{\PreserveBackslash\centering}m{\cellWidthe}>{\PreserveBackslash\centering}m{\cellWidthf}}
\toprule

\textbf{No.} & \textbf{Sample} & \textbf{Laser ($\bm{\uplambda}$\textsubscript{1}/$\bm{\uplambda}$\textsubscript{2}) in nm} & \textbf{Intensity in W/mm\textsuperscript{2}} & \textbf{Imaging Area in cm\textsuperscript{2}} & \textbf{Acquisition Times \textsuperscript{1}}\\
\cmidrule{1-6}

1 & Aspirin/Paracetamol & NIR (784.43/785.48) & 0.004 \textsuperscript{2} & 1 & 200 $\times$ 200 ms\\

2 & PS/PMMA & blue (457.74/458.90) & 0.1 & 0.02 & 40 $\times$ 100 ms\\

3 & Lipid-rich pork tissue & blue (457.74/458.90) & 0.2 & 0.02 & 40 $\times$ 50 ms and 1 $\times$ 2000 ms\\

4 & Pork tissue & blue (457.74/458.90) & 0.2 & 0.02 & 40 $\times$ 50 ms\\

\bottomrule
\end{tabular}}

    \begin{tabular}{c}
\multicolumn{1}{p{\linewidth-1.5cm}}{\footnotesize \justifyorcenter{\textsuperscript{1} Number of acquisitions x acquisition time for each excitation wavelength. Example: 40 $\times$ 50 ms results in a total acquisition time of 4 s for each excitation wavelength. The one-time final readout of the CCD takes 1 min. \textsuperscript{2}~The~micro lens array (MLA) facing the sample generates 400 $\times$ 0.25 mm\textsuperscript{2} laser foci with a power of approx. 1~mW~each. }}
\end{tabular}
    \end{table}

\subsection{Data Handling \label{sect:sec2dot4-sensors-974262}}

The raw data were extracted, the wavelength was calibrated using a mercury (neon) lamp and an argon lamp (PenRay, Quantum Design Europe, Germany), and intensity was calibrated by a broad emission white light source. All of these preprocessing steps have been performed using the software p3d~\cite{B54-sensors-974262,B55-sensors-974262,B56-sensors-974262}.~Afterwards, the data were further processed using the computer language R~\cite{B57-sensors-974262} and the following R packages: hyperSpec~\cite{B58-sensors-974262}, FITSio~\cite{B59-sensors-974262}, Ramancal~\cite{B60-sensors-974262}, pracma~\cite{B61-sensors-974262}, MALDIquant~\cite{B62-sensors-974262}, and~viridis~\cite{B63-sensors-974262}.

In a first step, all spectra were cosmic spike corrected as well as corrected for pixel errors. As~the MLAs were not exactly cut along the spaces in between the lenses, the MLAs were no more centrosymmetric.~One outer horizontal and outer vertical lens row moved close to border of the holder aperture, which caused some shading of these rows. Consequently, the top spaxel row and/or the last spaxel row to the right were not usable and discarded for some images. Then, in order to obtain a SERDS image, the spatially corresponding spectra of the two Raman images were pairwise subtracted from one another, always subtracting the higher excitation wavelength spectra from the lower excitation wavelength spectra.~A wavenumber calibration was performed on reconstructed spectra of paracetamol.~For Sample 1 and Sample 4, a hierarchical cluster analysis (HCA) was performed using the SERDS spectra as input. For this analysis, the Pearson correlation distance between the SERDS spectra was first calculated and then clustered following Ward’s minimum variance method~\cite{B64-sensors-974262}.~The reconstruction of the Raman spectra was calculated by summation of the signal intensities over the wavenumber channels. Since this reconstruction method leads to an underlying background in the reconstructed Raman spectrum, the sensitive nonlinear iterative peak (SNIP) algorithm was implemented.~The~SERDS spectra of Sample 4 were normalized by a vector normalization before clustering. For better visualization purposes, the difference spectra of sample 1 were not normalized. The normalization would apparently amplify the residual room light and would give a poor signal-to-noise~cluster.

\section{Results and Discussion \label{sect:sec3-sensors-974262}}
\unskip
\subsection{Wide-Field SERDS Imaging Using Nod and Shuffle and the Large Field of View (1 cm\textsuperscript{2}) Probe Head \label{sect:sec3dot1-sensors-974262}}
\unskip
\subsubsection{Illustration of the Raw Wide-Field Raman Images on the CCD Chip Using Nod and Shuffle \label{sect:sec3dot1dot1-sensors-974262}}

\begin{figure}[H]
\centering
\includegraphics[scale=2.2]{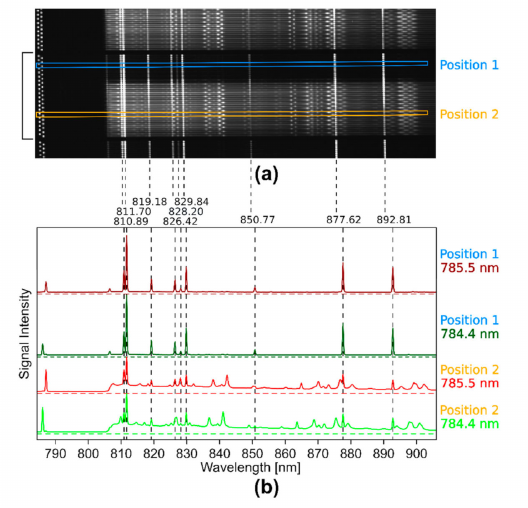}
\caption{Illustration of the raw, wide field Raman data measured with two different excitation wavelengths ($\uplambda$\textsubscript{1} = 784.4 nm, $\uplambda$\textsubscript{2} = 785.5 nm) and the nod and shuffle technique: (\textbf{\boldmath{a}}) zoomed view on a part of the CCD chip after acquisition for a closer look on the signal traces. Twenty spectra of each wavelength are always arranged in a block of traces (black square bracket), corresponding to one row of pixels in the Raman images. The upper blue box marks two intensity traces (Position 1) that correspond to a spaxel in the Raman image without a sample. The lower orange box marks two intensity traces (Position 2) that correspond to a spaxel on the Raman image with a paracetamol pill; (\textbf{\boldmath{b}})~signal intensities of the two signal traces (in dark red and dark green) marked by the upper blue box in the CCD image showing only the room light spectra. In red ($\uplambda$\textsubscript{1}) and green ($\uplambda$\textsubscript{2}), the signal intensities of the two signal traces marked by the lower orange box in the CCD image are shown, which~correspond to the paracetamol Raman spectra superposed by room light spectra. The dotted, vertical~black lines indicate the position of the fluorescent lamp bands and above their spectral position in~nm.}
\label{fig:sensors-974262-f004}
\end{figure}

To demonstrate the nod and shuffle technique, a Raman image of a paracetamol and an aspirin pill (sample 1) was measured with the large 1 cm\textsuperscript{2} field of view probe head (\fig{fig:sensors-974262-f004}a) configuration for 200 ms with 200 acquisitions before reading out the CCD camera. Since the laser power per fiber was low for a NIR laser source, this long acquisition time was necessary to obtain strong Raman signals. The two pills did not cover the probe head completely (\fig{fig:sensors-974262-f005}a). This was done on purpose to show that even room light directly shining into the detection system for a short period can be filtered out with this setup. One second after the measurement had started, the fluorescent lamp of room light was turned on for 3 s. In \fig{fig:sensors-974262-f004}a, a zoomed image of the CCD chip, which is turned by 90$^\circ$ compared to the schematic CCD chip in \fig{fig:sensors-974262-f002}, is shown for a better visualization. Higher signal intensity corresponds to higher brightness, and the horizontally arranged traces correspond to complete Raman spectra of two intertwined Raman images taken at two different excitation wavelengths $\uplambda$\textsubscript{1} and $\uplambda$\textsubscript{2}. The~traces are arranged in blocks, where each block consists of 20 spectra of each excitation wavelength corresponding to an image pixel row of 20 spaxels on the probe head. The black square bracket marks one of these blocks in \fig{fig:sensors-974262-f004}a.

The two Raman images are shifted on the CCD by 5 pixels in the vertical direction. Therefore, every second trace belongs to the same excitation wavelength and hence the Raman image and each neighboring wavelength pair belong to the same spatial position on the probe head. The oscillating dots on the left side of every trace correspond to the residual intensity of the excitation laser light that reached the detector in spite of the long-pass filter.~The spots on the right of the laser signals correspond to Raman bands that shift with the excitation wavelength and to very intense, non-shifting signals of the fluorescent lamp. \fig{fig:sensors-974262-f004}b shows the profile plots of the four signal traces ($\uplambda$\textsubscript{1} in green and dark green; $\uplambda$\textsubscript{2} in red and dark red) inside the blue (Position 1) and orange (Position 2) boxes in \fig{fig:sensors-974262-f004}a. At the position of the blue box, there are fluorescent lamp bands only, whereas at the position of the orange box, the Raman signals are superposed with these bands that are shown by black vertical, dotted lines. Their positions were indicated on top in nm and coincide well with the emission bands of Krypton~\cite{B65-sensors-974262}. After comparison of the profiles, the Raman bands can be discriminated from the fluorescent lamp bands. Some Raman bands overlap with the fluorescent lamp bands after the shift of excitation wavelength (red spectrum), which would make it hard to correct for the fluorescent lamp contributions with computational methods. According to the position of the Raman bands and their profile, the selected Raman spectra belong to paracetamol. The image of the CCD chip also shows how close together all the signal traces are. If the signal intensities on the exposed pixels on the CCD are too high, the intensities cause crosstalk in the surrounding pixels and even in the neighboring shifted spectra. This is a limitation of the nod and shuffle approach, requiring careful monitoring of the overall intensity distribution between the spectra and the use of appropriately weighted extraction schemes such as optimal extraction~\cite{B66-sensors-974262}.
%\unskip

\subsubsection{Filtering Out Room Light by SERDS Imaging \label{sect:sec3dot1dot2-sensors-974262}}

\begin{figure}[H]
\centering
\includegraphics[scale=1.6]{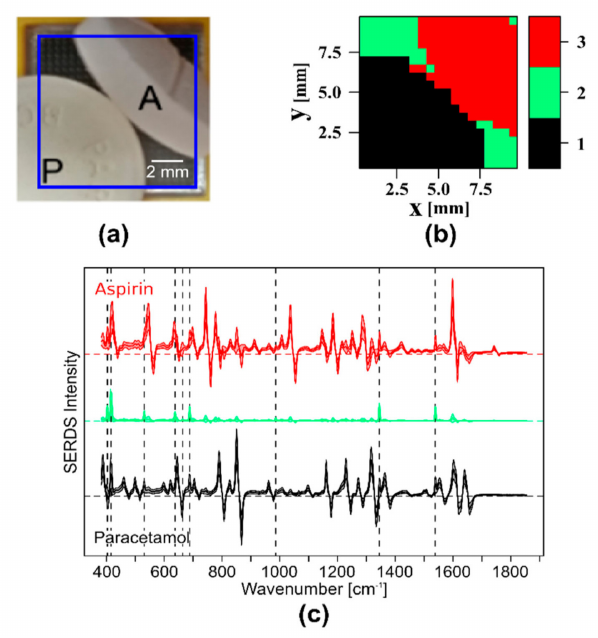}
\caption{SERDS imaging using nod and shuffle of a paracetamol (P) and an aspirin (A) pill with an acquisition of 200 times 200 ms. (\textbf{\boldmath{a}}) Picture of the sample on the probe head, the blue square corresponds to the field of view; (\textbf{\boldmath{b}}) cluster map of the SERDS image; (\textbf{\boldmath{c}}) clustered mean SERDS spectra and their standard deviation, cluster 1 = paracetamol (black), cluster 2 = no sample/residual room light (green), cluster 3 = aspirin (red); the dotted, vertical black lines indicate the position of the room light~bands.}
\label{fig:sensors-974262-f005}
\end{figure}

A strength of SERDS is the capability to filter out room light or ambient light with a distinct band structure. This has been shown for single spectra before~\cite{B67-sensors-974262,B68-sensors-974262,B69-sensors-974262} but not for a complete Raman image or, in our case, a wide field Raman image. Since the room light bands do not shift with the change in excitation wavelength, the difference spectra of the SERDS image are ideally free of any room light bands. This advantage is demonstrated with help of the aforementioned samples (\sect{sect:sec3dot1dot1-sensors-974262}) of paracetamol and aspirin. The room light illumination time of a fluorescent tube was limited to 3 s to avoid overexposure of the CCD and the above-mentioned signal crosstalk. As shown before in \fig{fig:sensors-974262-f004}b, the room light bands are very dominant in the Raman spectra of the two Raman images. After subtraction, the room light bands vanish almost completely (\fig{fig:sensors-974262-f005}c). An HCA cluster analysis of the SERDS image using three clusters, one for paracetamol, one for aspirin, and one for empty space as the probe head, was not completely covered by the two pills. The resulting cluster map is depicted in \fig{fig:sensors-974262-f005}b and agrees well with the original bright field image (\fig{fig:sensors-974262-f005}a). In \fig{fig:sensors-974262-f005}c, the mean SERDS spectra and their respective standard deviation of the three clusters are shown. These clusters can be identified based on their SERDS spectra as paracetamol (cluster 1, black), empty space with residual room light (cluster 2, green), and aspirin (cluster 3, red). Residual traces of room light are still visible, which might be due to uneven exposure of one of the Raman images resulting in a higher room light intensity as the other, since the room light had to be turned off to avoid overexposure. Still, the~strength of SERDS filtering out room light is obvious, which presents a huge advantage in operating theaters where room light needs to be switched on. Shorter acquisitions times will further reduce the interference of room light. However, the interface that was used to control a 784.43/785.48~nm laser source by polling the TTL (transistor-transistor-logic, square signal pulses that use +5  V for “1” or “high” and 0  V for “0” or “low”) trigger signals of the spectrograph with an USB digital input output device did not allow exposure times shorter than 200~ms.

\subsection{Wide-Field SERDS Imaging Using Nod and Shuffle, the Smaller Field of View Probe Head (0.02 cm\textsuperscript{2}), and~the Rapidly Shifting Dual-Wavelength Blue Diode Laser Source \label{sect:sec3dot2-sensors-974262}}
\unskip
\subsubsection{Wide Field SERDS Imaging with Nod and Shuffle to Differentiate Different Polymer Beads \label{sect:sec3dot2dot1-sensors-974262}}

\begin{figure}[H]
\centering
\includegraphics[scale=1.6]{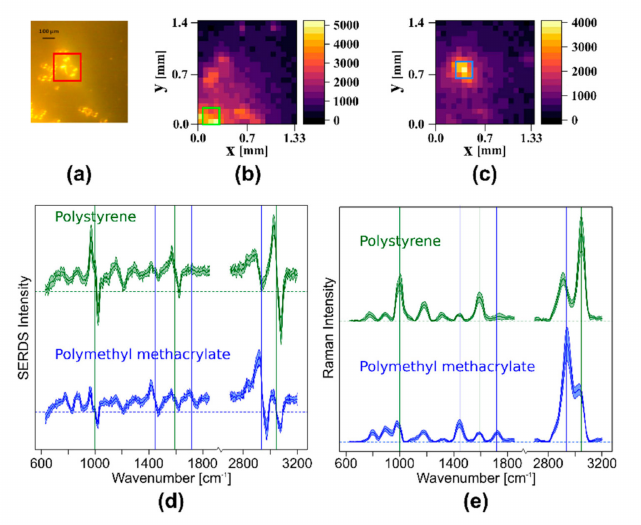}
\caption{SERDS imaging of polystyrene and polymethyl methacrylate beads: (\textbf{\boldmath{a}}) bright field image of the sample, red box shows a region of larger PMMA beads; (\textbf{\boldmath{b}}) PS intensity map based on the intensity of three marker bands in the baseline corrected, reconstructed Raman spectra; (\textbf{\boldmath{c}}) PMMA intensity map based on the intensity of three marker bands in the baseline corrected, reconstructed Raman spectra; (\textbf{\boldmath{d}}) mean SERDS spectra and their standard deviation for nine spectra with the highest intensities for PMMA and PS marked by boxes (green for PS, blue for PMMA) in the corresponding intensity maps; (\textbf{\boldmath{e}}) baseline corrected and reconstructed mean Raman spectra with standard deviation of the SERDS spectra shown in (\textbf{\boldmath{d}}); The vertical lines in (\textbf{\boldmath{d}},\textbf{\boldmath{e}}) indicate the marker bands (green for PS; blue for PMMA) that were used to create the intensity maps (\textbf{\boldmath{b}},\textbf{\boldmath{c}}).}
\label{fig:sensors-974262-f006}
\end{figure}

Since the spatial resolution of the larger field of view of 1 cm\textsuperscript{2} with 400 image points (1 spectrum per 0.5 mm) is quite low for polymer beads, the setup was changed to a smaller field of view of 0.02~cm\textsuperscript{2}, resulting in approximately 1 spectrum per 70 $\upmu$m. To demonstrate the imaging capability and the higher lateral resolution of this setup, polystyrene (PS) and polymethyl methacrylate (PMMA) beads were distributed on a CaF\textsubscript{2} slide and placed on the probe head. The Raman spectra were collected with the nod and shuffle technique with an acquisition time of 100 ms and 40 accumulations using the dual-wavelength blue diode laser light source ($\uplambda$\textsubscript{1} = 457.7 nm, $\uplambda$\textsubscript{2} = 458.9 nm). In the bright field camera image (\fig{fig:sensors-974262-f006}a), groups of small PS beads (50 $\upmu$m diameter) and a region of bigger PMMA beads (120 $\upmu$m diameter, red box) can be seen. To generate Raman intensity maps for PS and PMMA, specific marker bands of the reconstructed and baseline corrected Raman spectra were used. The PS intensity map (\fig{fig:sensors-974262-f006}b) shows the mean value of the aromatic ring breathing vibration 999 cm\textsuperscript{$-$1}, the ring stretching vibration at 1590 cm\textsuperscript{$-$1}, and the aromatic CH stretching vibration at 3047 cm\textsuperscript{$-$1}. To create the PMMA intensity map (\fig{fig:sensors-974262-f006}c), the mean value of the CH\textsubscript{3} deformation vibration at 1444 cm\textsuperscript{$-$1}, the C=O stretching vibration at 1718 cm\textsuperscript{$-$1}, and the CH stretching vibration in CH\textsubscript{2} and CH\textsubscript{3} at 2935 cm\textsuperscript{$-$1} was calculated and mapped. A comparison of the bright field image with the two Raman intensity maps of PS and PMMA shows that the position of the PMMA beads and the PS beads corresponds to the Raman intensity maps very well. The squares on the intensity maps show regions of high intensity in the respective bands (green box for PS, blue box for PMMA).

In \fig{fig:sensors-974262-f006}d, the mean SERDS spectra of these marked areas are shown in green (PS) and blue (PMMA). The vertical lines show the three marker bands for the intensity maps in the same colors. In \fig{fig:sensors-974262-f006}e, the mean and baseline corrected Raman spectra reconstructed from SERDS spectra of \fig{fig:sensors-974262-f006}d are plotted. The typical spectral profile of PS and PMMA are clearly visible, although~the bands are rather broad.~Since the laser was designed for the investigation of biological samples, the~large spectral shift of approximately 55 cm\textsuperscript{$-$1} is more appropriate for the SERDS investigation of tissue samples, which have a Raman profile with a larger band width. For the narrow-banded Raman profiles of polymers, this large shift results in the observed broadening during the integration in the reconstruction step and also yields a high risk of introducing artifacts into the reconstructed Raman spectra, since neighboring bands are shifted into one~another.

\subsubsection{Photobleaching Compensation in Wide Field SERDS Imaging by Very Fast Nod and Shuffle \label{sect:sec3dot2dot2-sensors-974262}}

A great challenge of SERDS is the variation in background intensities during the measurement, especially when the wavelength pair is affected differently. The main cause of changes in spectral background of biological samples is photobleaching of intrinsic fluorophores.~Their generated autofluorescence gets bleached by the continuous laser radiation, which mainly causes a uniform decrease of the fluorescence. This leads to a SERDS spectrum with a background slope that needs further correction with some kind of computational solution such as normalization or optimization. Another possibility to deal with this challenge is to use very short acquisitions, thereby only allowing for minimal changes. Here, the nod and shuffle technique with fast switching between the wavelengths is promising, since the accumulation of multiple cycles allows for enough Raman signal collection, while equally distributing the photobleaching effect on the Raman measurements at both excitation wavelengths.~In combining this with wide field imaging and hence simultaneous irradiation of all image pixels, differences within one SERDS image can be~eliminated.

Since the change in excitation wavelength was limited to 200 ms with the commercially available red laser system, the custom-built dual-wavelength blue diode laser system with the possibility for fast wavelength switching was used for the following experiments.~A change from a NIR to a blue excitation will usually lead to a huge increase of the amount of fluorescence especially for biological samples. However, this drawback should be eliminated by the SERDS~technique.

\begin{figure}[H]
\centering
\includegraphics[scale=.4]{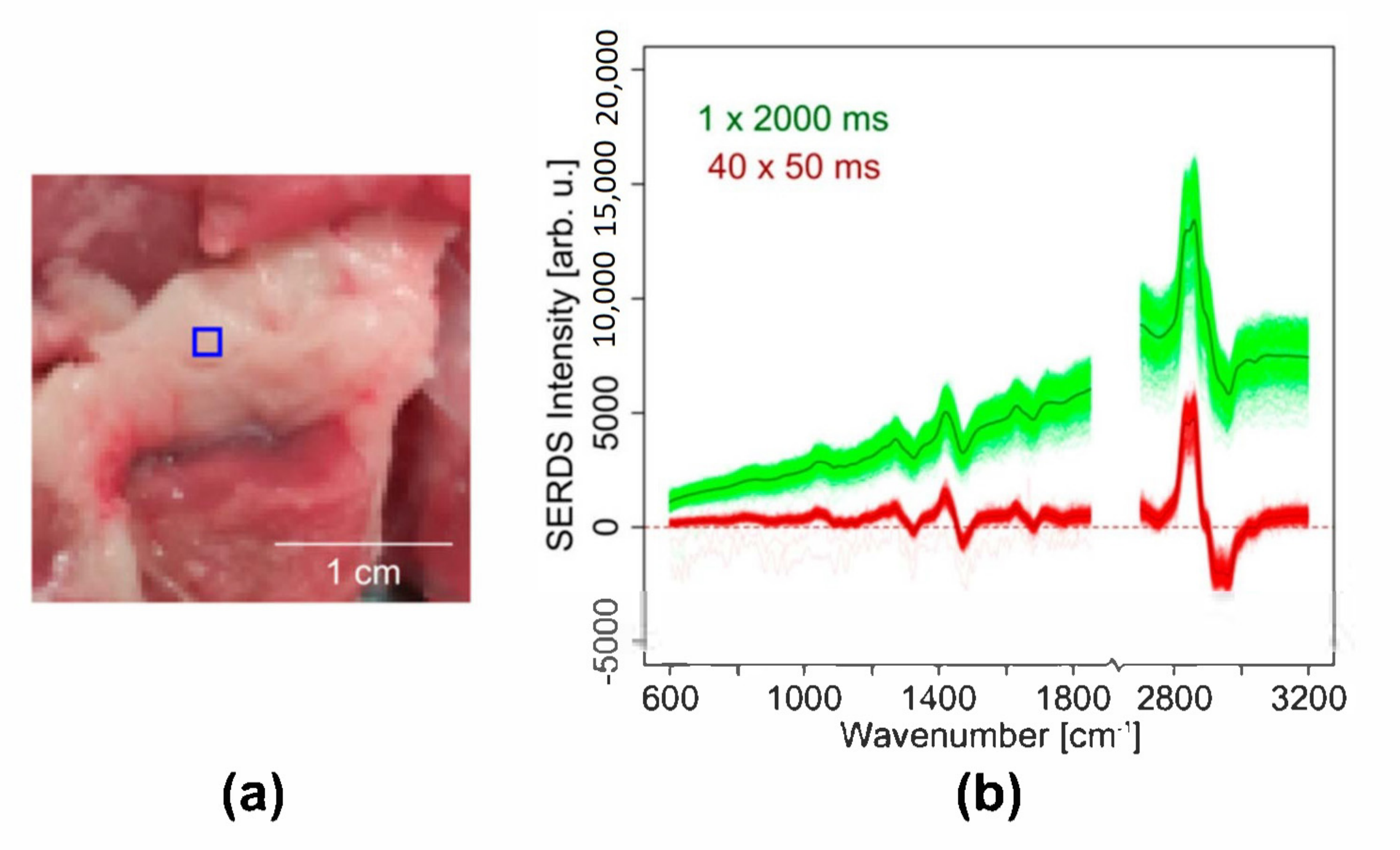}
\caption{Compensation of photobleaching in SERDS imaging of the fat region of a pork chop using nod and shuffle:~(\textbf{\boldmath{a}}) image of the pork chop sample with measurement area marked by a blue square; (\textbf{\boldmath{b}}) SERDS spectra of the SERDS image (red) with an acquisition time of 50 ms and 40~accumulations, SERDS spectra of the SERDS image (green) with an acquisition time of 2000 ms in a single accumulation. The darker lines (dark green and dark red, respectively) show the mean SERDS spectra of the respective SERDS image. The total exposure time is 4 s for each difference spectrum (2 s for each excitation~wavelength).}
\label{fig:sensors-974262-f007}
\end{figure}

To demonstrate the compensation of photobleaching in wide field SERDS imaging using nod and shuffle, a piece of pork meat (\fig{fig:sensors-974262-f007}a) was imaged ($\uplambda$\textsubscript{1} = 457.7 nm, $\uplambda$\textsubscript{2} = 458.9 nm) with a total acquisition time of 2 s in two approaches:\begin{enumerate}[label=\arabic*.]
\item  Two Raman images were measured with a 50 ms acquisition time and 40 accumulations for each excitation wavelength before reading out the CCD chip once. Thus, the overall acquisition time was 2 s for each excitation~wavelength.
\item  Two Raman images were measured with a 2 s acquisition time and one accumulation for each excitation wavelength before reading out the CCD chip once. Thus, the total acquisition time was also 2 s for each excitation~wavelength.
\end{enumerate}

The two resulting Raman images of each measurement were subtracted. The SERDS spectra for the two SERDS images, with the mean difference spectra marked as dark colored lines (see~\fig{fig:sensors-974262-f007}b), show~the typical difference profile of lipid-rich tissue with bands at 1446 cm\textsuperscript{$-$1}, 1650 cm\textsuperscript{$-$1}, and~1740~cm\textsuperscript{$-$1}. While the SERDS bands have comparable absolute intensities, it is clearly visible that the difference spectra show a stronger residual background for the longer acquisition time (green spectra) due to the aforementioned photobleaching effect: At the beginning of the measurement at $\uplambda$\textsubscript{1}, the pork shows remarkable autofluorescence that significantly decreases during the 2 s measurement time. At the beginning of the measurement at $\uplambda$\textsubscript{2}, the autofluorescence is already~decreased.

Thus, the two measurement pairs show different backgrounds, leading to a non-zero baseline after the subtraction.~For the short acquisition times, this residual background is not present, resulting~in difference spectra almost on the zero line (red spectra).~The reason for this effect is the equal distribution of the photobleaching on both measurements at $\uplambda$\textsubscript{1} and $\uplambda$\textsubscript{2}, when switching rapidly and repeatedly between the excitation wavelengths. Within a short 50 ms exposure time, the change of background signal is negligibly small. Each single measurement pair contains the same amount of background. When subtracting their sums, the background vanishes, and a zero baseline is obtained. This~clearly demonstrates the advantage of the nod and shuffle technique in combination with wide field SERDS imaging and rapid switchable laser diodes. Even though the photobleaching must have been higher in the image with the short acquisition times, since it was recorded first, this setup configuration successfully circumvents this challenge and distributes the photobleaching equally on both excitation~wavelengths.

\subsubsection{Wide Field SERDS Imaging of Different Tissues in Pork Meat Using Nod and Shuffle \label{sect:sec3dot2dot3-sensors-974262}}

\begin{figure}[H]
\centering
\includegraphics[scale=1.6]{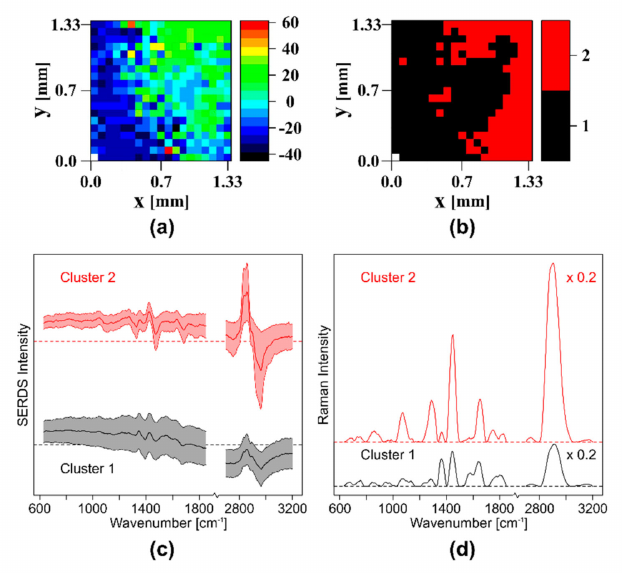}
\caption{SERDS imaging of a piece of pork meat with protein and lipid-rich areas: (\textbf{\boldmath{a}}) intensity map of the SERDS spectra at 2900 cm\textsuperscript{$-$1}; (\textbf{\boldmath{b}}) cluster map of the SERDS spectra; (\textbf{\boldmath{c}}) clustered normalized, mean~SERDS spectra and their standard deviation; (\textbf{\boldmath{d}}) reconstructed and baseline corrected mean Raman spectra of the two clusters shown in (\textbf{\boldmath{c}}). The high wavenumber regions were multiplied by 0.2~for better~visualization.}
\label{fig:sensors-974262-f008}
\end{figure}

To test the introduced wide field SERDS setup on a heterogeneous biological sample, another~piece of pork meat was imaged.~As before, the custom made dual-wavelength blue diode laser source (\mbox{$\uplambda$\textsubscript{1} = 457.7 nm}, $\uplambda$\textsubscript{2} = 458.9 nm) and the smaller field of view probe head with the nod and shuffle technique was used to image a section including proteins and lipids. \fig{fig:sensors-974262-f008}a shows an intensity map of the SERDS spectra based on the intensity at 2900 cm\textsuperscript{$-$1}.

After hierarchical clustering, two clusters visualize the regions with protein and lipids in the cluster map (see \fig{fig:sensors-974262-f008}b).~In \fig{fig:sensors-974262-f008}c, the normalized, mean difference spectra with standard deviation of the two clusters are shown. It is obvious that the intensity of cluster 2, the lipid cluster, is much higher than the intensity of cluster 1, the protein cluster. The larger standard deviation is a consequence of the lower intensity and the higher background in the protein cluster. For a better visualization of the band positions, the reconstructed and baseline corrected mean Raman spectra of the two clusters can be seen in \fig{fig:sensors-974262-f008}d. Typical protein bands such as amide I at 1638 cm\textsuperscript{$-$1}, amide~III at 1282 cm\textsuperscript{$-$1}, myoglobin at 1363 cm\textsuperscript{$-$1}~\cite{B32-sensors-974262}, tyrosine at 858 cm\textsuperscript{$-$1}, tryptophane (754,~894~cm\textsuperscript{$-$1} and 1587~cm\textsuperscript{$-$1}), and CN (1075 cm\textsuperscript{$-$1}, 1099 cm\textsuperscript{$-$1} and 1128 cm\textsuperscript{$-$1}) classify cluster 1 as proteins, while~cluster 2 shows lipid typical bands (carbonyl at 1750 cm\textsuperscript{$-$1}, C=C 1649 cm\textsuperscript{$-$1}, CH\textsubscript{2} and CH\textsubscript{3} at 1060 cm\textsuperscript{$-$1}, 1300~cm\textsuperscript{$-$1} and 1450 cm\textsuperscript{$-$1}) and can therefore be assigned to lipids. This also explains the already observed intensity differences between the two clusters, since lipids have a higher Raman scattering cross-section than proteins. In conclusion, with the chosen setup configuration, SERDS images of a biological sample can be recorded in one shot with small photobleaching effects, and the lipid and protein-rich regions can be easily~identified.

\section{Conclusions \label{sect:sec4-sensors-974262}}

Using wide field Raman imaging, several spectra of an area can be recorded in a relatively short time, since all spectra are collected simultaneously. Consequently, the total acquisition time of a Raman image, in our case consisting of 400 spectra of a 20 by 20 fiber array, is orders of magnitude faster than in point-by-point Raman imaging. The fibers were arranged in a square probe head to image an area of 10 $\times$ 10 mm\textsuperscript{2} with 0.5 $\times$ 0.5 mm\textsuperscript{2} per pixel in standard mode or 1.4 $\times$ 1.4 mm\textsuperscript{2} with \mbox{0.07 $\times$ 0.07 mm\textsuperscript{2}} per pixel in higher magnification mode. As the readout times of the usually employed, very large CCD chips are quite long, measurements in quick successions cannot be realized in this configuration. This is especially problematic for wide field SERDS imaging, where the accumulation of short measurements at two slightly different excitation wavelengths is~favorable.

The presented wide field SERDS imaging approach using the nod and shuffle technique allows accumulating very short acquisition times before reading out the CCD chip. The structure of the signal traces on the CCD chip was presented.~Since the signal traces on the CCD chip are closer together in the interlaced nod and shuffle approach, the risk of crosstalk from very high intensities into the neighboring spectra belonging to the other excitation wavelength can be problematic. For this reason, the signal intensity has to be monitored closely during the~experiments.

Using the nod and shuffle approach, the advantages of SERDS and fast acquisition times was demonstrated for wide field SERDS imaging for the first time. As expected from previous single spectra measurements, room light filtering was one of these advantages, where the measurement was interfered by an external light source during the acquisition. This was compensated due to the SERDS technique and the fast acquisition times, which made it possible to filter out the room light, even when it directly shone into the detection system for a short period. If the sample covers the probe head completely, only~stray light can enter the detection system. This will allow much longer acquisition times. The~other demonstrated advantage of the fast acquisition times is the compensation of photobleaching resulting in minimal background in the SERDS spectra of the SERDS images.~This~compensation makes the usual normalization or optimization steps before the calculation of the SERDS spectra unnecessary and the whole SERDS approach faster and less~computational.

It was demonstrated that structural differences can be resolved based on almost background free wide field SERDS images. The resulting reconstructed Raman spectra had to be baseline corrected, since the used quasi integration introduced a background into the reconstructed spectra. Especially for the investigation of the narrow band Raman profiles of polymers, the 1.16 nm wavelength shift of the custom-made, dual-wavelength blue diode laser, which corresponds to a 55 cm\textsuperscript{$-$1} shift of the Raman spectra, as well as the spectral broadening due to the reconstruction was quite large. For~the investigation of biological samples, which was the targeted application for the blue diode laser, this~large spectral shift is more~appropriate.

The nod and shuffle technique will enable a compensation of the photobleaching and by directly using the difference spectra for a classification of the measured samples, no reconstruction will be necessary.~This combination will streamline the wide field SERDS imaging and reduce the computational effort for this instrumental background correction~method.

\section{Patents \label{sect:sec5-sensors-974262}}

For results of the work, a patent application was filed at the German Patent and Trade Mark Office. Patent application no. DE 10 2018 130 582~A1.

\vspace{6pt}
%\newpage
\authorcontributions{Conceptualization: M.M.R., E.S., C.K., J.P. Realizing of the setup: E.S. Performing the measurements: F.K., E.S. Upgrading of the data reduction software: C.S. (Christer Sandin). Data analysis and visualization: F.K., T.U., G.M. Evaluation and validation: C.S. (Clara Stiebing), C.K. Design and realization of dual wavelength (457.74 nm/458.90 nm) diode laser source: A.M., M.M., B.S., G.T. Supervision: C.S. (Clara Stiebing), C.K., J.P. Original draft preparation: F.K. Review and editing of the manuscript: all. All authors have read and agreed to the published version of the~manuscript.}
\funding{This research was funded by the Leibniz Association through the project HYPERAM (SAW-2016-IPHT-2). The publication of this article was funded by the Open Access Fund of the Leibniz~Association.}
\acknowledgments{This research has made use of the integral-field spectroscopy data-reduction software p3d, which is provided by the Leibniz-Institute for Astrophysics Potsdam (AIP) and maintained by Sandin Advanced Visualization (Stockholm). We thank Thomas Fechner from the AIP’s electronics workshop for upgrading the spectrograph. Furthermore, we thank DECHEMA e.V. for the possibility to present results at the 14. Dresdner Sensor Symposium. Parts of the figures are adapted from the conference proceeding~\cite{B32-sensors-974262}.

}
\conflictsofinterest{The authors declare no conflict of interest. The funders had no role in the design of the study; in the collection, analyses, or interpretation of data; in the writing of the manuscript, or in the decision to publish the~results.}
\reftitle{References}

\end{document}